\newif\ifpreprint
  \preprintfalse	
  \preprinttrue		

\newif\ifastroph
  \astrophfalse		
  \astrophtrue		

\ifpreprint
	\documentstyle[aas2pp4,epsf,twoside,fleqn]{article}
	\newcommand{\fL}[1]{{\footnotesize\bf\uppercase{#1}}}
\else
 	\documentstyle[12pt,aasms4,epsf]{article}
\fi

\newcommand{\bea}	{\begin{array}}
\newcommand{\eea}	{\end{array}}
\newcommand{\beq}	{\begin{equation}}
\newcommand{\eeq}	{\end{equation}}
\newcommand{\ben}	{\begin{eqnarray}}
\newcommand{\een}	{\end{eqnarray}}
\newcommand{\bsq}	{\begin{mathletters}}
\newcommand{\esq}	{\end{mathletters}}

\newcommand{\ds}	{\displaystyle}
\newcommand{\B}[1]	{\mbox{\boldmath$#1$}}
\newcommand{\bx}	{\B{x}}
\newcommand{\bvel}	{\B{v}}
\newcommand{\bw}	{\B{w}}
\newcommand{\D}		{{\rm d}}
\newcommand{\op}	{\omega_\phi}
\newcommand{\oR}	{\omega_R}
\newcommand{\sR}	{\sigma_R}
\newcommand{\Rcr}	{R_{\rm\scriptscriptstyle CR}}
\newcommand{\Rolr}	{R_{\rm\scriptscriptstyle OLR}}
\newcommand{\Ro}	{R_0}
\newcommand{\Rb}	{R_{\rm b}}
\newcommand{\vc}	{v_{\rm c}}
\newcommand{\vo}	{v_0}
\newcommand{\Ab}	{A_{\rm b}}
\newcommand{\Af}	{A_{\rm f}}
\newcommand{\volr}	{v_{\rm\scriptscriptstyle OLR}}
\newcommand{\Ob}	{\Omega_{\rm b}}
\newcommand{\Oo}	{\Omega_0}
\newcommand{\fuv}	{\mbox{$f_0(u,v)$}}
\newcommand{\half}	{\case{1}{2}}

\newcommand{\SgrA}	{Sgr\,A$^{\!\star}$}
\newcommand{\kpc}	{\mbox{\,kpc}}
\newcommand{\mkms}	{\mbox{km\,s$^{-1}$}}
\newcommand{\kms}	{\mbox{\,km\,s$^{-1}$}}
\newcommand{\kmskpc}	{\mbox{\,km\,s$^{-1}$\,kpc$^{-1}$}}
\newcommand{\mkmskpc}	{\mbox{km\,s$^{-1}$\,kpc$^{-1}$}}
\newcommand{\etal}	{\mbox{\em et~al.}}

\newcommand{\eqn}[1]	{equation\ (\ref{#1})}
\newcommand{\eqi}[1]	{equation\ [\ref{#1}]}
\newcommand{\eqb}[1]	{(\ref{#1})}
\newcommand{\Sec}[1]	{Section\ \ref{sec:#1}}
\newcommand{\Fig}[1]	{Figure\ \ref{fig:#1}}
\newcommand{\fig}[1]	{Fig.\ \ref{fig:#1}}
\newcommand{\figs}[2]	{Figs.\ \ref{fig:#1} and \ref{fig:#2}}
\newcommand{\Tab}[1]	{Table\ \ref{tab:#1}}

\newcommand{\amp}	{\ifpreprint{}\else\&\fi}

\ifpreprint
\else
  \lefthead{Walter Dehnen}
  \righthead{Effect of the Galactic Bar on the Local Velocity Distribution}
\fi

\begin{document}

\ifpreprint \thispagestyle{empty} \fi
\title{The Effect of the Outer Lindblad Resonance of the Galactic Bar\\
		 on the Local Stellar Velocity Distribution}
\author{Walter Dehnen}
\affil{	Theoretical Physics, 1 Keble Road, Oxford OX1 3NP, United Kingdom and \\
	Max-Planck Institut f.\ Astronomie, K\"onigstuhl, 69117 Heidelberg,
	Germany; dehnen@mpia-hd.mpg.de}

\begin{abstract} \ifpreprint\noindent\fi
Hydro-dynamical modeling of the inner Galaxy suggest that the radius of the
outer Lindblad resonance (OLR) of the Galactic bar lies in the vicinity of the
Sun. How does this resonance affect the distribution function in the outer parts
of a barred disk, and can we identify any effect of the resonance in the
velocity distribution actually observed in the solar neighborhood? To answer
these questions, detailed simulations of the velocity distribution, $f(\bvel)$,
in the outer parts of an exponential stellar disk with nearly flat rotation
curve and a rotating central bar have been performed. For a model resembling the
old stellar disk, the OLR causes a distinct feature in $f(\bvel)$ over a
significant fraction of the outer disk. For positions up to 2\kpc\ outside the
OLR radius and at bar angles of $\sim$10-70 degrees, this feature takes the form
of a bi-modality between the dominant mode of low-velocity stars centred on the
local standard of rest (LSR) and a secondary mode of stars predominantly moving
outward and rotating more slowly than the LSR.

Such a bi-modality is indeed present in $f(\bvel)$ inferred from the Hipparcos
data for late-type stars in the solar neighborhood. If one interpretes this
observed bi-modality as induced by the OLR -- and there are hardly any viable
alternatives -- then one is forced to deduce that the OLR radius is slightly
smaller than $\Ro$. Moreover, by a quantitative comparison of the observed with
the simulated distributions one finds that the pattern speed of the bar is
$1.85\pm0.15$ times the local circular frequency, where the error is dominated
by the uncertainty in bar angle and local circular speed.

Also other, less prominent but still significant, features in the observed
$f(\bvel)$ resemble properties of the simulated velocity distributions, in
particular a ripple caused by orbits trapped in the outer 1:1 resonance.
\end{abstract}


\ifpreprint		\section{I\fL{ntroduction}} \label{sec:intro} \noindent
\else			\section{Introduction} \label{sec:intro}
\fi
The fact that the Milky Way is barred has been established first of all by the
interpretation of the gas velocities observed in the inner Galaxy (\cite{vau64};
\cite{pet75}; \cite{coh76}; \cite{lb80}; \cite{gv86}; \cite{ml86}; \cite{b91})
and later confirmed by infrared photometry (\cite{bs91}; \cite{wei94}; 
\cite{dw95}; \cite{bgs97}) and asymmetries in the distribution or magnitude of
stars (\cite{nak91}; \cite{wc92}; \cite{wb92}; \cite{nw97}; \cite{st95};
\cite{sev96}; \cite{st97}). However, there is still substantial debate on the
structure and morphology of the bar, its orientation with respect to the Sun, and
the rotation rate, usually referred to as pattern speed. Recent hydro-dynamical
investigations of Englmaier \& Gerhard (1999) and Weiner \& Sellwood (1999) and
the combined stellar- and gas-dynamical models of Fux (1999a) suggest that the bar
rotates fast, i.e.\ that corotation occurs somewhere between 3.5 and 5\kpc\
(for $\Ro=8\kpc$) not much beyond the end of the bar. Moreover, the bar angle,
the azimuth of the Sun with respect to the bar's major axis, is restricted both
by photometry and kinematics to lie in the range between about $10^\circ$ and
$45^\circ$.

Apart from the corotation resonance (CR), a rotating bi-symmetric bar creates
two other fundamental resonances, the  inner (ILR) and outer (OLR) Lindblad
resonances, which occur when
\beq \label{reso}
	\Ob = \op \mp \half \oR.
\eeq
Here, $\op$ and $\oR$ are the azimuthal and radial orbital frequencies, while
$\Ob$ is the pattern speed of the bar. An orbit for which relation \eqb{reso} is
satisfied is in phase with the bar, i.e.\ after each completion of one radial
lobe the orbit is at the same position with respect to the bar. Thus, a star
on a resonant orbit is always pushed in the same direction, and thus forced 
into another orbit.

For circular orbits, $\op=\Omega(R)$ (circular frequency) and $\oR=\kappa(R)$
(epicycle frequency) and each resonance can be mapped uniquely to the radius of
the resonant circular orbit; hereafter $\Rolr$ and $\Rcr$ denote the radii where
circular orbits are in outer Lindblad and corotation resonance. For a nearly
flat Galactic rotation curve, $\Rolr\approx1.7\Rcr$ and, from the aforementioned
estimates for $\Rcr$, we find $\Rolr\sim6$-9\kpc. Thus, we expect the OLR of the
Galactic bar to be located in our immediate Galactic surrounding. The goal of
this paper is to answer the question whether and how this proximity of the
resonance affects the stellar kinematics observable in the solar neighborhood.

\subsection*{Stellar Dynamics Near the OLR} \label{sec:olr}
\ifpreprint \noindent \fi
Using linear perturbation theory for near-circular orbits (cf.\ Binney \&
Tremaine 1987, p.~146-151), one finds that closed orbits in a barred potential
are elongated either parallel or perpendicular to the bar. The orientation
changes at each of the fundamental resonances: inside ILR orbits are anti-aligned
(so-called x$_2$ orbits), between the ILR and CR they are aligned (x$_1$
orbits), while between CR and OLR the orbits are anti-aligned, until they align
again beyond the OLR. 

\placefigure{fig:olr}
\ifpreprint
  \begin{figure}[t]	 
	\centerline{\epsfxsize=55mm\epsfbox[90 220 550 676]{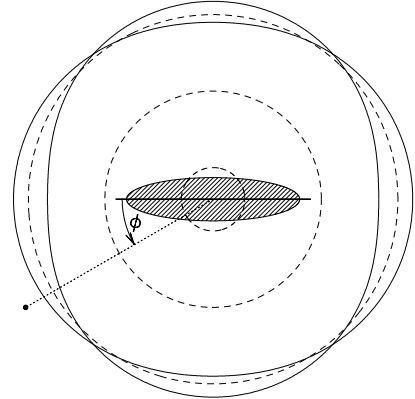}}
	\caption[]{\footnotesize 
	Closed orbits ({\it solid\/}) just inside and outside the OLR of a
	rotating central bar ({\it shaded ellipse\/}). The circles ({\it
	dotted\/}) depict the positions of the ILR, CR, and OLR (from inside out)
	for circular orbits. Note the change of the orbits' orientation at the
	OLR, resulting in the crossing of closed orbits at four azimuths.  A
	possible position of the Sun is shown as filled circle. The bar angle
	$\phi$ is indicated for the case of a clockwise rotating bar.
	\label{fig:olr} }
  \end{figure}
\fi

The situation at the OLR is sketched in \Fig{olr}, which shows two closed orbits
(solid curves) just in- and outside the OLR as they appear in a frame of
reference corotating with the bar. In this frame, the orbits near OLR rotate
counter-clockwise for a clockwise rotating bar, like that of the Milky
Way. Thus, at bar angles $\phi$ between $0^\circ$ and $90^\circ$, the closed
orbits inside OLR move slightly outwards, while those outside OLR move inwards.
Clearly, if all disk stars moved on closed orbits, the stellar kinematics would
deviate from that of a non-barred galaxy only at positions very close to
$\Rolr$, where the closed orbits are significantly non-circular. In particular,
at azimuths where the closed orbits from either side of the OLR cross, one would
expect two stellar streams, one moving inwards and the other outwards%
\footnote{
	Based on this consideration, Kalnajs (1991) suggested that the Hyades
	and Sirius stellar streams in the solar neighborhood are caused by the
	Sun being located almost exactly at one of the two possible positions in
	the Galaxy where such orbit crossing occurs with the appropriate sign of
	the radial velocity difference.}.

However, in general stellar orbits are not closed, but exhibit radial
oscillations. Many of these orbits are trapped into resonance (\cite{wb94}), and
may be described, to lowest order in their eccentricity, as epicyclic
oscillations around a closed parent orbit. This means in particular, that such
trapped eccentric orbits from inside the OLR, i.e.\ with $\Ob<\op+\half\oR$, can
visit locations outside $\Rolr$, and vice versa. Thus, the stellar velocity
distribution observable in the solar neighborhood, $f_0(\bvel)$, shall be
affected by the OLR whenever near-resonant orbits pass near the Sun {\em and\/}
are sufficiently populated. We may estimate, using the epicycle approximation in
a back-of-the-envelope calculation, that for late-type disk stars this condition
is satisfied for $\Rolr$ in the range from%
\footnote{
	The {\sc rms} epicycle amplitude of stars in a population with radial
	velocity dispersion $\sR$ is $X\approx\sqrt{2}\,\sR/\kappa$. Stars
	originating from a radius $R=\Ro+x$ could visit us, if $|x|<X(R)$ yielding
	\[
		|x| \la	 \sqrt{2}\;{\sR(\Ro)\over\kappa(\Ro)} \,
		\left[1-x\left(R_0^{-1}+R_\sigma^{-1}\right)+{\cal O}(x^2)\right],
	\]
	where I have assumed that $\sR$ decays exponentially with scale length
	$R_\sigma$ and that $\kappa\propto R^{-1}$ (as for flat rotation
	curves). With $\sR(\Ro)\approx35\kms$, $\kappa(\Ro)\approx35\kmskpc$,
	and $\Ro=R_\sigma=8\kpc$, this results in the two solutions $x\approx-2$
	and 1\kpc.}
6 to 9\kpc\ (for $\Ro=8\kpc$). This coincides with the above estimate for
$\Rolr$. Thus, we expect the OLR of the Galactic bar to affect the velocity
distribution observable locally for late-type stars. Based on the properties of
the closed orbits, one would expect also for stars moving on non-closed orbits
different typical velocities depending whether they originate from in- or
outside the OLR. Thus, the expected effect on $f_0(\bvel)$ is a bi-modality.

Weinberg (1994) has studied the orbital response to a rotating bar using
epicycle theory and assuming a flat rotation curve. He found indeed that
different orbital families may overlap creating a bi-modality in the velocity
distribution (though he only considered the resulting increase in the velocity
dispersions). Weinberg also considered a slowly decreasing pattern speed, which
affects the relative number of stars trapped into resonance with the bar, but
leaves the final phase-space position of the resonance, and hence the velocity of
possible modes in $f_0(\bvel)$, unchanged.

In order to verify and quantify these estimates and expectations of the response
of a warm stellar disk to a stirring bar and its OLR, I performed numerical
simulations, presented in \Sec{simul}. In \Sec{orbs}, I discuss the closed
orbits in the outer part of a barred disk galaxy and their relation to the
features apparent in the local $f(\bvel)$, while \Sec{offset} gives quantative
estimates for the velocity of the secondary mode induced by the OLR. In
\Sec{comp}, the velocity distribution observed in the solar neighborhood, which
indeed shows a bi-modality, is compared both qualitatively and quantitatively to
those emerging from the simulations. Finally, \Sec{conc} concludes and sums up.

Throughout this paper, I will use units of kpc and \mkms\ for radii and
velocities, while frequencies and proper motions are given in \mkmskpc.
\ifpreprint		\section{T\fL{he} S\fL{imulations}} \label{sec:simul}
			\noindent
\else			\section{The Simulations} \label{sec:simul}
\fi
In order to investigate the typical structure of a dynamically warm stellar disk
in the presence of a rotating bar, we simulate the slow growth of a central bar
with constant rotation frequency in an initially axisymmetric equilibrium
representing a warm stellar disk. Clearly, this does not simulate the true
evolution of the Galaxy, since the stellar disk was hardly dynamically warm
already when the bar formed. Moreover, the bar has presumably developed too,
both in strength and in pattern speed. However, we are not aiming at simulating
the formation of the Milky Way, but at answering the question for the effect
of the OLR, which significantly alters the phase-space structure, independently
of the formation history of the Galaxy.

\subsection{The Simulation Technique} \label{sec:simul:tech}
\ifpreprint \noindent \fi
Currently, traditional $N$-body simulations are not suitable for studying the
influence of the central bar on a stellar disk at high resolution: the required
number of particles exceeds $10^8$ already for merely two-dimensional
simulations, which makes them very CPU-time intensive and aggravates any
investigation of the parameter space. However, $N$-body simulations are
unnecessary luxury, as they provide dynamical self-consistency and model the
whole system, both of which is not required in this study. Instead, we (i)
assume the growth of a central bar, rather than simulate a bar-instability
self-consistently, and (ii) only integrate trajectories which are crucial to the
local velocity distribution, rather than those for the whole system. Moreover,
Poisson noise is avoided by placing the trajectories on a regular grid in the
{\em observables} and integrate backward in time (see \Sec{integ} for details).

This technique, which may be called {\em backward integrating restricted
$N$-body method}, is similar in spirit to that used in the method of perturbation
particles (\cite{lcb93}), and has first been used by Vauterin \& Dejonghe (1997,
1998), who studied the non-linear evolution of the stellar distribution function
(DF) in the inner parts of a bar-unstable model, whereby using the analytic
potential due to the dominant linear bar mode.

Because we are only interested in the effects on the planar motions in the outer
disk, our modeling is two-dimensional, and we do not care much about the details
of the inner Galaxy.
\subsubsection{The Initial Equilibrium} \ifpreprint \noindent \fi
Since we are interested only in near-circular trajectories passing through a
point in the outer parts of the disk, a possible inadequateness of our model
potential in describing the inner parts of the Galaxy is unimportant.
Therefore, we take a simple power-law rotation curve
\bsq \label{monopole}
\beq \label{vc}
	\vc = \vo\,(R/\Ro)^\beta
\eeq
as our initial axisymmetric model. Here, $\Ro$ denotes the Sun's distance from
the Galactic center and $\vo$ the local circular speed. The corresponding
underlying potential,
\beq \label{Phi}
        \Phi_0(R) = \vo^2 \times \left\{ \bea{l@{$\quad$}l}
        \ds (2\beta)^{-1}\,(R/\Ro)^{2\beta}	& {\rm for\;}\beta\neq0\\[1ex]
        \ds \ln(R/\Ro)				& {\rm for\;}\beta=0,
\eea \right. \eeq \esq
is meant to originate partly from the gaseous and stellar components (disk, 
bulge/bar, halo) and partly from dark matter. In restricting our model to two 
dimensions, we make no assumptions about the vertical distribution of the matter
generating the potential \eqb{Phi}.

The dynamical equilibrium of the stellar disk is described by a simple DF,
$f=F(E,L_z)$, which depends on energy $E$ and angular momentum $L_z$ of the
stars, and is given by equation (10) of Dehnen (1999b). The parameters of the DF
are set such that it generates, to very good approximation, an exponential disk
with scale length $R_s$ and an exponential velocity-dispersion profile with
scale length $R_\sigma$ and local radial velocity dispersion
$\sigma_0\equiv\sR(\Ro)$.

\subsubsection{The Bar Potential} \ifpreprint \noindent \fi
The contribution of the bar to the potential in the outer disk is dominated
by the quadrupole\footnote{
	Bar formation is mainly a re-arrangement of the matter inside 
	corotation such that the monopole of the potential outside this 
	region remains unchanged.},
and we neglect higher-order multipoles for simplicity. Some orbits visiting
the outer disk may have passed through the inner disk and bar region. Therefore,
we will use a slightly more elaborate model for the bar potential than that 
resulting from just a constant quadrupole moment:
\beq
	\Phi_{\rm b} = \Ab\,\cos\left(2[\phi-\Ob t]\right)\,
		\times\left\{ \bea{l@{$\quad$}l}
		-(\Rb/R)^3	& {\rm if}\;R\ge\Rb, \\[2ex]
		(R/\Rb)^3-2	& {\rm if}\;R\le\Rb, \eea \right.
\eeq
where $\Rb$ and $\Ob$ are the size and pattern speed of the bar. The amplitude 
$\Ab$ of the quadrupole is switched on smoothly. It is zero before $t=0$,
grows with time at $0<t<t_1$ as
\beq \label{ampl}
	\Ab = \Af \left({3\over16} \xi^5 - {5\over8} \xi^3 
			+ {15\over16} \xi + {1\over2}\right),
	\quad \xi \equiv 2 {t\over t_1} - 1,
\eeq
and stays constant at $\Ab=\Af$ after $t_1$. Thus, the amplitude and its 
first and second time derivative behave continuously for all $t$, guaranteeing 
a smooth transition from the non-barred to the barred state.

\subsubsection{The Time Integration} \label{sec:integ} \ifpreprint \noindent \fi
The collisionless Boltzmann equation tells us that the DF remains constant along
stellar trajectories. Thus, the value $f(\bw,t_2)$ of the DF at some phase-space
point $\bw\equiv(\bx,\bvel)$ and time $t_2$ is equal to $f(\bw_0,0)$ if $\bw$
originates from integrating $\bw_0$ from $t=0$ to $t=t_2$. In our case and in
contrast to $N$-body simulations, the potential as function of space and time is
known a priori. Therefore, we can obtain $\bw_0$ just by integrating the orbit
passing through $\bw$ at $t_2$ backward in time until $t=0$.

In practice, we specify the final time $t_2$ and choose the phase-space points
$\bw$ from a grid in planar velocities at a position $(\Ro,\phi)$ in the
Galactic plane. Each of the resulting orbits is integrated backward until
$t=0$, and the initial energy $E$ and angular momentum $L_z$ are
remembered. From these, the value $f(\bw,t_2)=f(\bw_0,0)\equiv F(E,L_z)$ can be
computed for any initially axisymmetric equilibrium DF $F$.
\subsection{The Parameters} \ifpreprint \noindent \fi
Some of the parameters arising in the simulations are unimportant for our
purposes. For instance, the size $\Rb$ of the bar, which, in agreement with the
findings mentioned in \Sec{intro}, we fix to be 80\% of the corotation radius.
For our model, the latter is given by
\beq \label{Rcr}
	\Rcr \equiv\Ro\,(\Ob/\Oo)^{1/(\beta-1)},
\eeq
where $\Oo\equiv\vo/\Ro$ denotes the local circular frequency.

Since our simulations are not self-consistent, the parameters of the DF are
unimportant for the {\em dynamics\/} of the stellar orbits, but crucial for
their {\em population\/} with stars. In this study, we will use a DF designed to
resemble the old stellar disk. The exponential surface density has scale length
$R_s=0.33\Ro$, the exponential velocity dispersion is normalized to
$\sR(\Ro)=0.2\vo$ and has scale length $R_\sigma=\Ro$. I also experimented
with $R_\sigma=0.66\Ro$, corresponding to $\sR^2\propto\Sigma$, and found
very similar results.
\subsubsection{The Time Scales} \ifpreprint \noindent \fi
It turns out that the time $t_1$, determining how smoothly the bar is switched 
on, is not very important for the outcome of the simulations. However, the total
integration time under influence of the bar, i.e.\ $\approx t_2-t_1/2$, has a 
significant impact on the {\em details\/} of the resulting velocity 
distributions. We will use $t_2=2t_1$ and consider various values for
$t_1$ in units of the bar rotation period $T_{\rm b}\equiv2\pi/\Ob$.
\subsubsection{The Solar Position} \ifpreprint \noindent \fi
The relative distance of the Sun from the bar is conviently parametrized by the
ratio $\Rolr/\Ro$, where for the power-law models \eqb{monopole}
\beq \label{Rolr}
	\Rolr = \Ro \left({\Oo\over\Ob}\,
		\left[1+\sqrt{1+\beta\over2}\right]\right)^{1/(1-\beta)}.
\eeq
Another important parameter is the bar angle $\phi$, the angle by which the Sun
is behind the bar's major axis (cf.\ \fig{olr}).
\subsubsection{The Rotation Curve} \ifpreprint \noindent \fi
The shape of the rotation curve, parametrized by the power $\beta$ (see
\eqi{vc}), dictates the energy dependence of the orbital frequencies. We use a
flat rotation curve ($\beta=0$) as default but will also consider slightly
rising or falling rotation curves.

Changing the local circular speed $\vo$, or equivalently the frequency $\Oo$, 
at fixed $\Rolr/\Ro$ corresponds to a change of the pattern speed $\Ob$ by the 
same factor. Such a change is essentially a re-scaling of the model and does 
not require any additional simulations.
\subsubsection{The Bar Strength} \ifpreprint \noindent \fi
The strength of the bar is best measured in a way that is independent of such
re-scaling. We will use as parameter the dimensionless quantity
\beq \label{alpha}
	\alpha \equiv 3\,{A_{\rm f}\over\vo^2}\,\left({\Rb\over\Ro}\right)^3,
\eeq
which is the ratio of the forces due to the bar's quadru\-pole and the 
axisymmetric power-law background at galactocentric radius $\Ro$ on the bar's 
major axis.
\subsection{Scanning the Parameter Space} \ifpreprint \noindent \fi
Simulations have been performed for $\Rolr/\Ro$ between 0.8 and 1.2 at steps of
0.05, nine values of $\beta$ at steps of $0.05$ between $-0.2$ and $0.2$, and
bar angles $\phi$ at $5^\circ$ intervals between $10^\circ$ and $50^\circ$.
Additionally, the integration time and bar strength have been varied. In the
figures below, unless stated otherwise, all the parameters but one are fixed to
the default values listed in \Tab{default}, in order to show the effect of one
parameter alone. These default parameters correspond to an OLR less than 1\kpc\
inside the solar circle, a bar angle of 25$^\circ$ as discussed from studies of
the gas motions, and a flat rotation curve.

In order to ease a comparison with the observations, the velocity distributions
arising from the simulations are displayed in the $u$ and $v$, defined,
respectively, as the velocity towards the Galactic center and in direction of
rotation, both relative to the circular orbit passing through the position of
the Sun in absence of any bar.

\placetable{tab:default}
\ifpreprint
  \begin{figure}
	\ifastroph
		\centerline{\epsfxsize=83.54mm\epsfbox[142 -226 496 666]{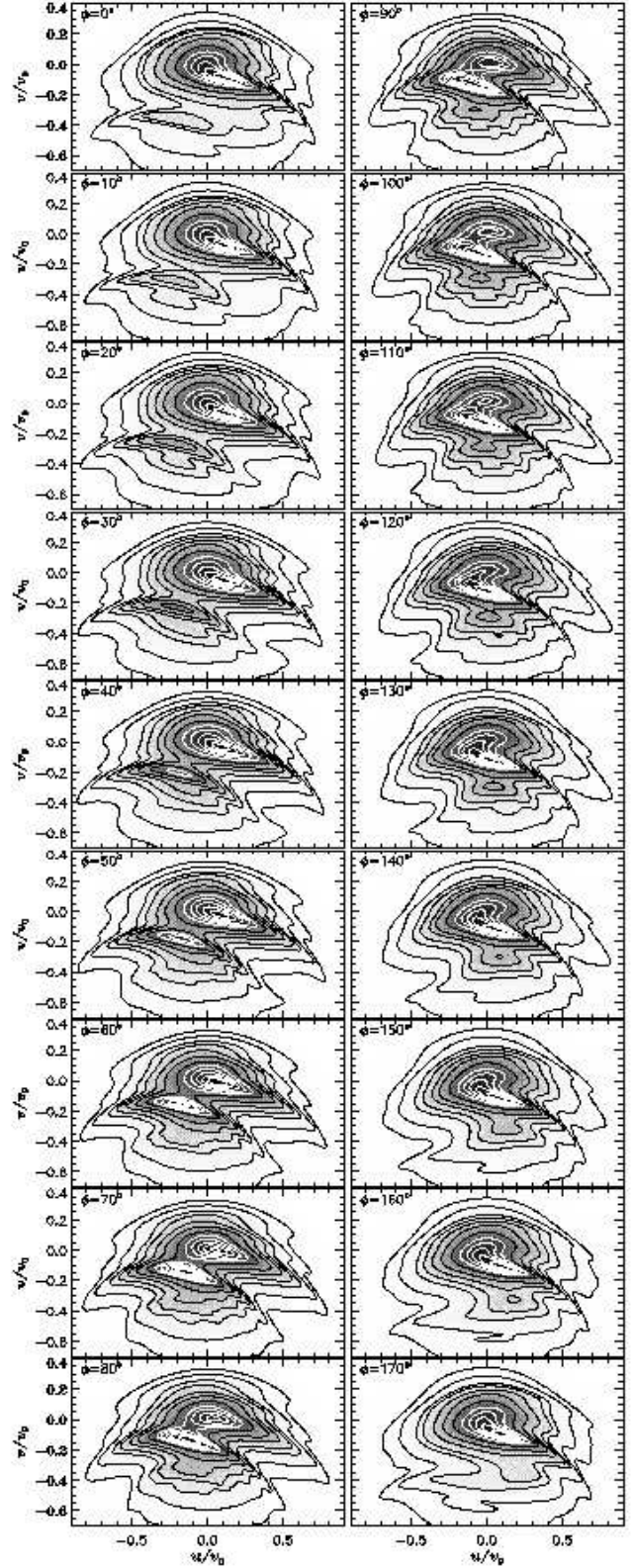}}
	\else
		\centerline{\epsfxsize=83.54mm\epsfbox[6 4 252 634]{Dehnen.FIG2.ps}}
	\fi

	\caption[]{\footnotesize 
	Simulated velocity distributions \fuv: variation with bar angle.
	The remaining parameters are fixed at their default values in
	\Tab{default}. \label{fig:varphi} }
  \end{figure}
  \begin{figure}[t]
	\centerline{\epsfxsize=60mm\epsfbox[7 6 169 274]{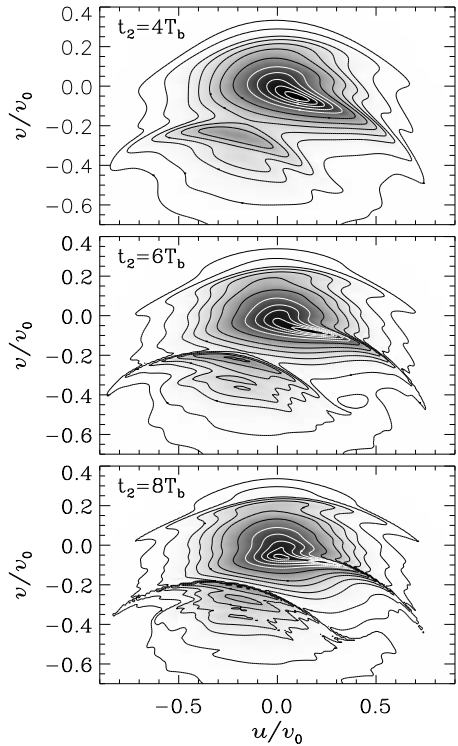}}
	\caption[]{\footnotesize 
	Simulated velocity distributions \fuv: variation with 
	integration time $t_2$ ($T_b$ denotes the bar rotation period).
	The bar-growth time is set to $t_1=0.5t_2$.
	\label{fig:vartime} }
  \end{figure}
  \begin{figure}[t]
	\centerline{\epsfxsize=60mm\epsfbox[6 19 159 411]{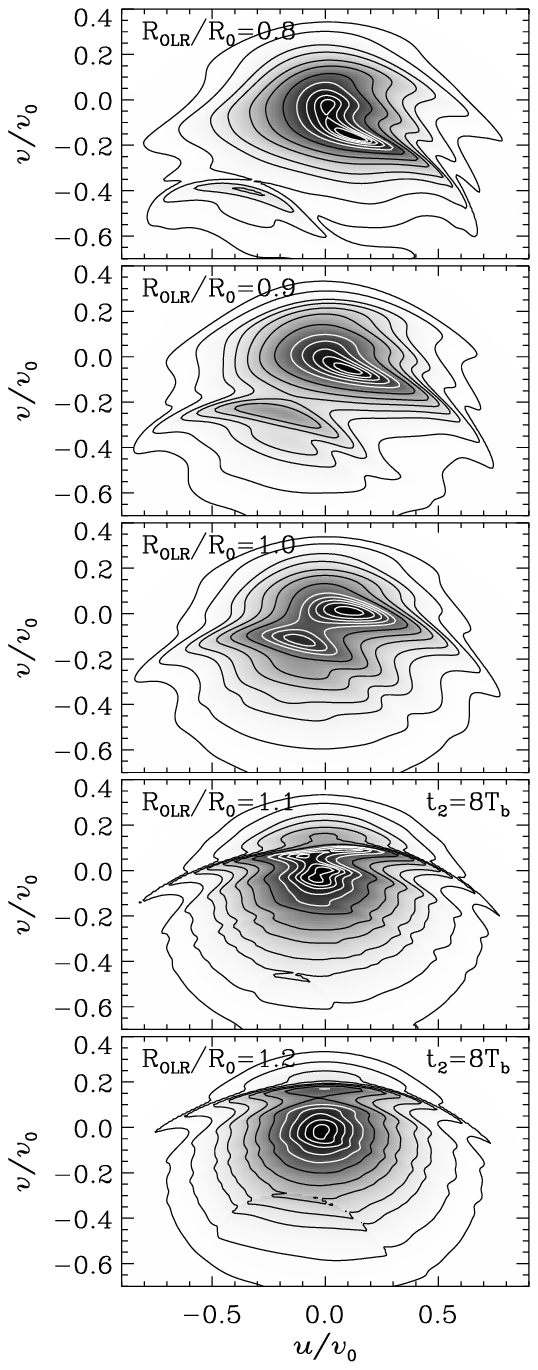}}
	\caption[]{\footnotesize 
	Simulated velocity distributions \fuv: variation with $\Rolr/\Ro$. 
	\label{fig:varpos} }
  \end{figure}
  \begin{figure}[t]
	\centerline{\epsfxsize=60mm\epsfbox[7 0 166 425]{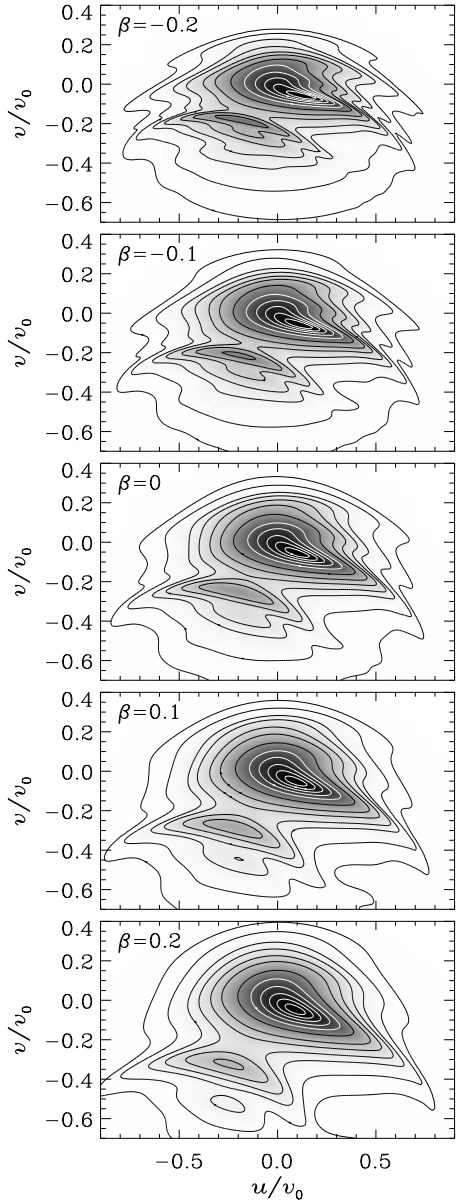}}
	\caption[]{\footnotesize 
	Simulated velocity distributions \fuv: variation with $\beta$,
	parameterizing the shape of the rotation curve. \label{fig:varbeta} }
  \end{figure}
  \begin{figure}[t]
	\centerline{\epsfxsize=60mm\epsfbox[7 6 169 274]{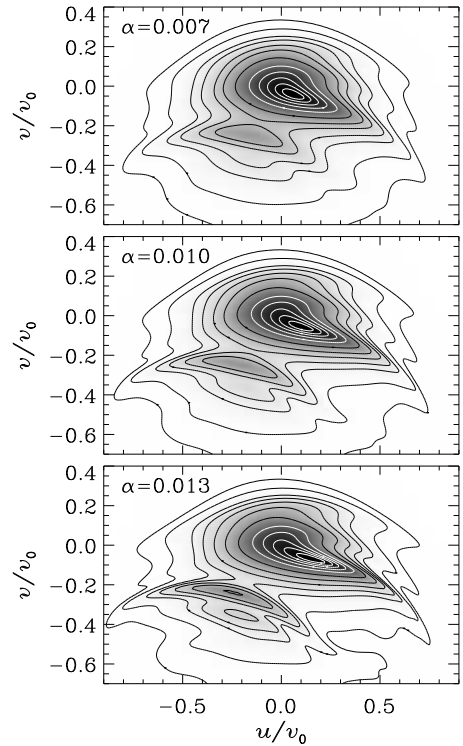}}
	\caption[]{\footnotesize 
	Simulated velocity distributions \fuv: variation with bar 
	strength, parameterized by $\alpha$. \label{fig:varalpha} }
  \end{figure}
  \begin{table}[t]
	\begin{center} \refstepcounter{table} \label{tab:default}
			Table~\thetable \\
		Default values for the simulation parameters
	\begin{tabular}{llr@{}c@{}l}
								\\[-2ex] 
								\hline\hline
								\\[-2.3ex]
		\multicolumn{2}{c}{parameters to be varied} & 
		\multicolumn{3}{c}{default value}		\\ \hline
								\\[-2ex]
	shape of rotation curve & $\beta$	&0 &		 &	\\
	bar angle 		& $\phi$	&25&&$\!\!{}^\circ$	\\
	position w.r.t.\ OLR	& $\Rolr/\Ro$	&0 &.	&9		\\
	bar strength		& $\alpha$	&0 &.	&01		\\
	integration time 	& $t_2$		&4 &  &$T_{\rm b}$	\\ 
	\hline \\[-2.3ex]
	\multicolumn{2}{c}{parameters kept fixed}	& 
	\multicolumn{3}{c}{default value}				\\ 
	\hline \\[-2ex]
	bar size		& $\Rb/\Rcr$	& 0  &. &8		\\
	disk scale length	& $R_s/\Ro$	& 0  &. &33		\\
	local velocity dispersion & $\sR(\Ro)/\vo$& 0&. &2		\\
	$\sR$ scale length	& $R_\sigma/\Ro$& 1  &	&		\\
	bar growth time		& $t_1$		& 0  &. &5 $t_2$	\\ 
	\hline\hline
	\end{tabular}
	\end{center}
  \end{table}
\fi 

\subsubsection{Variations with Bar Angle} 
\ifpreprint 	\noindent 
\else		\placefigure{fig:varphi} \fi
\Fig{varphi} plots \fuv\ at intervals of $10^\circ$ in $\phi$ (note that the 
situation is bi-symmetric, so that $\phi$ and $\phi+180^\circ$ are identical)
after the growth of a central bar. A bi-modality is clearly visible for bar
angles in the first quadrant. As judged from these simulations, the {\em
existence\/} of the bi-modality in $f_0$ seems not to depend critically on the
bar angle. 

When increasing the bar angle, the depression between the two modes, hereafter
`LSR mode' (at $\bvel\sim0$) and `OLR mode' (at $u,v<0$), is shifted towards
higher $v$, while the OLR mode, which consists of stars originating from inside
the OLR, becomes more prominent. However, the latter effect is hard to compare
quantitatively with the observed \fuv, because the relative strength of the
modes is likely to be affected both by the parameters of the DF and changes of
the bar's pattern speed during the past (\cite{wb94}). In $u$, the OLR mode
ranges from about $-0.6$ to 0.2 in units of $\vo$.

There is also a clear distortion of the LSR mode, which, for bar angles relevant
to our position in the Milky Way, has the form of an extension to positive $u$
velocities at $v$ comparable to that of the OLR mode.
\subsubsection{Variations with Integration Time}
\ifpreprint 	\noindent 
\else		\placefigure{fig:vartime} \fi
\Fig{vartime} plots \fuv\ for three different choices of $t_2$ (with 
$t_1=0.5t_2$). Obviously, the longer integration times lead to more prominent
resonant features, not only due to the OLR but also the resonance $\Ob=\op+\oR$,
which is responsible for the ripple at $\vo\approx0.2$. Clearly, the longer the
action of non-axisymmetric forces, the stronger their imprints and the more
detailed the resulting structures in stellar velocities. Other effects of longer
integration times are a slight shift of the peak $v$ velocity of the OLR mode
towards lower values and a change in the structure of the LSR mode. However, the
velocity of the minimum between the two modes hardly changes, justifying the use
of $t_2=4T_b$ as default.

These simulations, assuming a constant pattern speed and a completely flat
rotation curve without any substructures like spiral arms, will certainly create
much more such fine resonant features in \fuv\ than the more noisy and less
constant force field of a real galaxy would produce. However, the locus of the
resonance itself is not such a detail and we can safely assume that the
$v$-velocity of the division line between the modes is not subject to artefacts
due to the idealizations made.
\subsubsection{Variations with Radius}
\ifpreprint 	\noindent 
\else		\placefigure{fig:varpos} \fi
\Fig{varpos} plots \fuv\ for five different choices of the observers distance
from the OLR, where for $\Rolr/\Ro=1.1$ and 1.2, the integration time is taken
to be $t_1=8T_b$ in order to strengthen any possible resonant features. The
bi-modality is clearly visible at $0.8\ga\Rolr/\Ro\ga1.05$ -- note, however,
that this `visibility' depends on the parameters of the DF and the bar angle
$\phi$, cf.~\fig{varphi}.

For $\Rolr/\Ro\ga1.05$, the OLR does not create a clear bi-modality (at least
not at $\phi=25^\circ$) but a distortion in \fuv\ in form of a ripple at roughly
constant energy (curves of constant energy in the $uv$ plane are circles centred
on $u=0$, $v=-\vo$). For $\Rolr/\Ro=1.2$, most stars are on orbits between
corotation and outer Lindblad resonance. Despite this proximity to two
fundamental resonances, the resulting \fuv\ is remarkably regular: apart from
the ripple caused by the OLR, it forms a nice elliptical distribution (bottom
panel of \fig{varpos}).
\subsubsection{Variations with Rotation-Curve Shape}
\ifpreprint 	\noindent 
\else		\placefigure{fig:varbeta} \fi
\Fig{varbeta} plots \fuv\ for five different choices of $\beta$, which
determines the shape of the rotation curve \eqb{vc}. Changing $\beta$ affects
the orbital frequencies, in the sense that they fall off more steeply with
energy for a falling rotation curve ($\beta<0$) than for a rising one
($\beta>0$). For a fixed difference in frequencies, a steeper fall-off results
in a smaller energy difference, which at fixed position yields a smaller
velocity distance. Therefore, reducing $\beta$ squashes the structures of the
velocity distribution in $v$: the extent in $v$ of the OLR mode shrinks and its
peak $v$ velocity becomes less negative (the opposite happens when increasing
$\beta$). Similarly, the distance in $v$ between the OLR and higher-order
resonances (responsible for the ripples at $v>0$) decreases with decreasing
$\beta$
\subsubsection{Variations with Bar Strength}
\ifpreprint 	\noindent 
\else		\placefigure{fig:varalpha} \fi
\Fig{varalpha} plots \fuv\ for three different bar strengths, parameterized 
by the force ratio $\alpha$ \eqi{alpha}. Apparently, stronger bars lead to
more pronounced OLR (as well as other resonant) features in \fuv: the 
OLR mode contains more stars and extends over a larger range in $u$ velocity. 

The $v$ velocity dividing the LSR and OLR modes is hardly changed. This is not
what one would naively expect from linear theory, in which the amplitude of the
orbital changes is proportional to the amplitude of the perturbation. However,
large-amplitude orbital changes are not suitably described by linear theory, but
result, in the case of the OLR, in orbits that are detuned by the non-linearity
and thus leave the near-resonant region of phase space. Thus, the OLR mode is
made from orbits which are shifted by similar amplitudes largely independent of
the strength of the bar perturbation.
\ifpreprint
  \begin{figure}[t]
	\centerline{\epsfxsize=60mm\epsfbox[28 166 410 706]{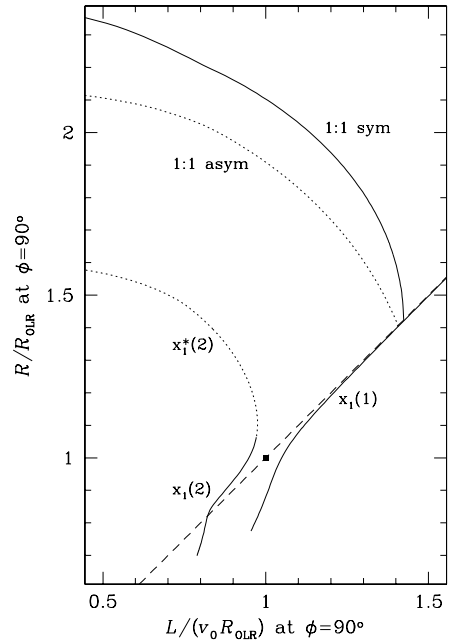}}
	\caption[]{\footnotesize
	Characteristic diagram (plot of $R$ vs.\ $L$ on the bar's minor axis) of
	the stable ({\em solid}) and unstable ({\em dotted}) closed orbits in
	the outer parts of the model with default settings. Orbits corresponding
	to high-order resonances ($n<-2$) are not displayed. The dashed line
	corresponds to the circular orbits in the unbarred model with the square
	indicating the OLR. \label{fig:char}}
  \end{figure}
  \begin{figure}[t]
	\centerline{\epsfxsize=85mm\epsfbox[238 204 560 688]{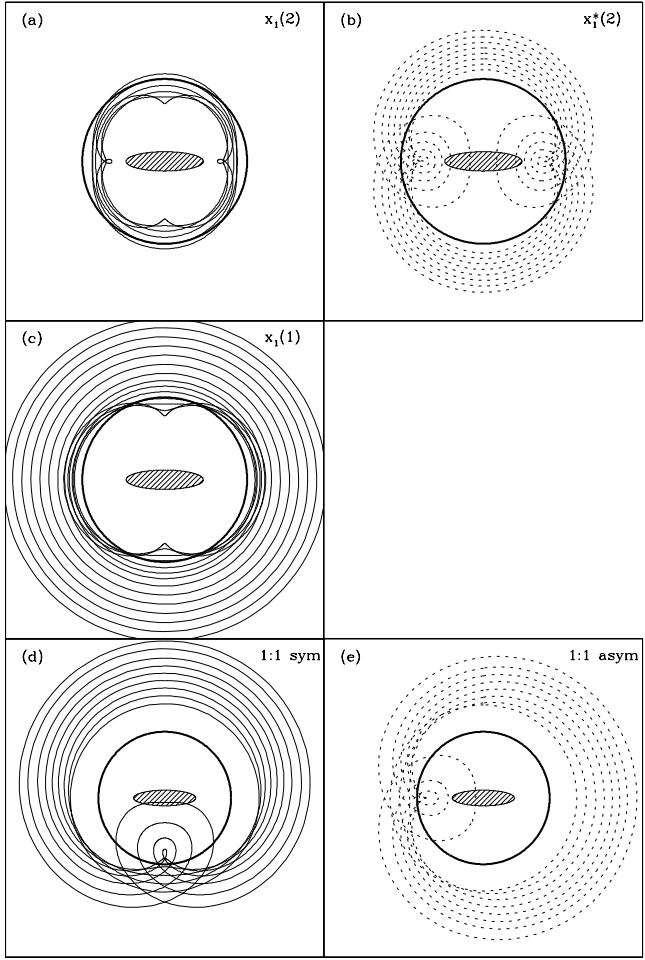}}
	\caption[]{\footnotesize
	Closed orbits of the families also displayed in \fig{char}.  The OLR is
	indicated by the bold circle. Solid and dotted curves refer to stable
	and unstable orbits, respectively.  The x$_1$(1) orbits extend to larger
	radii, while for the other four orbital families the innermost and
	outermost orbit is shown (note the difference in scale). The 1:1
	resonant orbits (d \& e) also exist as reflected versions, such that
	each family as a whole is bi-symmetric.  \label{fig:orbits}}
  \end{figure}
\fi

\ifpreprint	\section{R\fL{elation to} C\fL{losed} O\fL{rbits}}
		\label{sec:orbs}\noindent
\else		\section{Relation to Closed Orbits}\label{sec:orbs}
\fi
When a bar is slowly grown, the initially circular orbits will be mapped into
(nearly) closed, but no longer circular, orbits in the barred potential. These
are also the orbits on which gas is supposed to move, since encounters are
avoided. Thus, we expect most stars in a barred galaxy to move on nearly closed
orbits. Consequently, the properties of the velocity distributions are better
understood in light of the properties of the closed orbits that exist in the
outer parts of a barred disk.
\subsection{Closed Orbits in the Outer Disks of Barred Galaxy}
\ifpreprint\noindent\fi
The only literature on this topic appears to be a review by Contopoulos \&
Grosb{\o}l (1989), who consider closed orbits in the inner and outer parts of
disks with bars or spiral structure.

Outside of corotation, the main family of closed orbits are the near-circular
x$_1$ orbits. At each integer value of the rotation number
\beq \label{rot}
	n = {\oR\over\op-\Ob},
\eeq
however, this family is modified by a resonance. At resonances with even $n$ (as
e.g.\ the OLR, which has $n=-2$), this modification takes the form of a gap with
a distinct change in properties, such as shape, between the orbits on either
side of the resonance. At resonances with odd $n$, two families of orbits, one
symmetric and one anti-symmetric with respect to the bar's minor axis, branch
off the main x$_1$ family. Immediately outside corotation, the denominator in
\eqn{rot} scans through all small negative numbers, resulting in a whole host of
high-order resonances with $-\infty<n<-2$. \Fig{char} shows the characteristics
of the orbit families with $n\ge-2$ in the model with the default parameters of
\Sec{simul}.

\placefigure{fig:char}
\placefigure{fig:orbits}

At $n=-2$ (OLR) we have the typical change, already mentioned in the
introduction, from orbits inside OLR, which anti-align with the bar and are
called x$_1$(2) orbits by Contopoulos \& Grosb{\o}l, to orbits outside OLR,
called x$_1$(1) and aligning with the bar. The first subfamiliy, x$_1$(2),
extends to unstable orbits, called x$_1^\star$(2), at larger radii. Some
orbits of these families are shown in the upper three panels of \Fig{orbits}.
The innermost x$_1$(2) orbit clearly shows the effect of the $n=-4$ resonance.

The resonance with $n=-1$, also called --1:1 or outer 1:1 resonance, creates a
pair of orbit families (\fig{orbits}\,d\&e), each of which exists in two
versions, which are reflection symmetric to each other. The family which is
symmetric with respect to the bar's minor axis consists of stable orbits, while
the antisymmetric family contains only unstable orbits. At large radii, both
these family develop inner loops, which may penetrate right into the central
bar.
\subsection{Closed Orbits and \boldmath$f_0$}
\ifpreprint\noindent\fi
All of the orbital subfamilies in \Fig{orbits} pass through positions with
$R\sim\Rolr$, but there is no position $(R,\phi)$ which is visited by orbits
from all five families. Thus, the accessibility by closed and nearly closed
orbits changes between different positions in the outer Galaxy. Since each
stable family (and non-closed orbits trapped around it) will create a distinct
feature in the velocity distribution, we expect that \fuv\ changes almost
discontinuously between different positions, depending on the closed orbits
families passing through that position.
\subsubsection{The Position $\Rolr/\Ro=0.9$ and $\phi=25^\circ$}
\ifpreprint\noindent\fi
Let us firts consider the case of the simulation with default parameters, cf.\
the middle panels of \figs{varbeta}{varalpha}. There are three closed orbits
reaching this position: two x$_1$(1) orbits and one x$_1^\star$(2) orbit.

The first of the x$_1$(1) orbits has $(u,v)=(0.02,-0.015)\vo$, while for the
second $(u,v)=(0.09,-0.1)\vo$. This latter orbit originates from very close to
the OLR, resulting in a large pertubation, which enables its visit at
$\Ro$. Both these velocities lie within the LSR mode, but while the first is
very close to the undisturbed circular orbit, the local velocity of the latter
is related to the extension towards $u>0$. Thus, we can understand these
extensions of the LSR modes as caused by stars on nearly closed, hence strongly
populated, and nearly OL-resonant, hence highly disturbed, orbits.

The third orbit reaching $(\Ro,\phi)$ is unstable and has
$(u,v)=(-0.38,-0.19)\vo$. This velocity is exactly in the valley between the two
modes. Thus, it seems that this valley is caused by unstable, i.e.\ chaotic,
orbits.
\subsubsection{Varying the Radius}
\ifpreprint\noindent\fi
Let us now consider what happens if we change the distance from the OLR. For
$0.7\la\Rolr/\Ro\la1$, there is an unstable x$_1^\star$(2) orbit passing
through, and we therefore expect a depression (a valley or a truncation) in
$f_0$.

For $\Rolr/\Ro\ga1$ (the detailed value depends on $\phi$), there is a
fundamental change, as then (i) no stable closed x$_1$(1) orbits passes through,
but (ii) a stable x$_1$(2) instead of an unstable x$_1^\star$(2) orbit. These
two changes are certainly related to the rather abrupt change of the morphology
of \fuv\ apparent in \Fig{varpos}, from a clear bi-modality to a mere density
enhancement at roughly constant energy.
\subsubsection{Varying the Azimuth}
\ifpreprint\noindent\fi
If we change the bar angle $\phi$ at fixed $\Rolr/\Ro$, the families of closed
orbit passing through $(\Ro,\phi)$ change as well. In particular for
$\Rolr/\Ro\sim1$, the stable x$_1$(2) orbits appear for $\phi\ga40^\circ$, which
in \fig{varphi} leads to a stronger OLR mode for $\phi\sim90^\circ$ than for
$\phi\sim0^\circ$.

On the other hand, when increasing $\phi$, the second stable x$_1$(1) orbits
disappear with the consequence that the extension of the LSR modes towards $u>0$
weakens as one moves from the bar's major axis to its minor axis in
\fig{varphi}.
\subsubsection{The Outer 1:1 Resonance}
\ifpreprint\noindent\fi
Even though the stable 1:1 resonant closed orbits do not pass through positions
near the OLR with $|\phi|\la40^\circ$, there is a clear resonant feature in the
velocity distributions at these positions. Hence, this feature, which takes the
form of a density enhancement at certain velocities, must be due to related
non-closed resonant orbits. The density enhancement is likely caused by the
intersection of the various orbits of this family at
$0^\circ\la\phi\la180^\circ$.
\ifpreprint	\section{Q\fL{uantivying the} OLR F\fL{eature}}
		\label{sec:offset} \noindent
\else		\section{Quantivying the OLR Feature} \label{sec:offset}
\fi
In order to allow for a quantitative comparison of the OLR feature in the
simulated \fuv\ with the observations in the next section, we need to use a
well-defined quantity that is easy to measure and likely to be stable under
deviations from the idealized conditions assumed in the simulations. Here, I
will only use the $v$-velocity of the saddle point between the LSR and OLR mode.
This velocity, hereafter denoted $\volr$, is essentially independent of the
details of the stellar DF and even the strength of the bar.
\subsection{An Axisymmetric Estimate} \ifpreprint\noindent\fi
We may first neglect the effect of the bar potential and estimate the velocities
(at $\Ro$) of the orbits that would become exactly resonant in the presence of 
a bar with pattern speed $\Ob$. For power-law potentials, the orbital 
frequencies are well approximated by the corresponding frequencies of the 
circular orbit with the same energy (\cite{deh99a}). Using this approximation 
and the relations for power-law potentials (cf.\ Appendix B of Dehnen 1999a) 
and neglecting terms ${\cal O}(\bvel^3/\vo^3)$, we may estimate that the local 
velocities of orbits that are in OLR satisfy
\beq \label{v-olr}
	v +{u^2\over2\vo} \cong \tilde\volr \equiv \vo\,{1+\beta\over1-\beta} 
	\left[1-{\Ob/\Oo\over1+\sqrt{(1+\beta)/2}} \right].
\eeq
Thus, the local velocities of resonant orbits form a parabola whose maximal $v$
occurs at $u=0$ and depends on the distance from the OLR and the shape of the
rotation curve. Note that the valleys between the two modes in the simulated
\fuv\ are actually nearly parabolic. They are, however, displaced from the $u=0$
axis and also in $v$. These displacements must be due to the quadrupole forces
of the bar, neglected in the above estimate.
\subsection{Quantifying the Local OLR Velocity } \ifpreprint\noindent\fi
The value for the velocity $\volr$ is sensitive mainly to four parameters: the 
bar angle $\phi$, the relative distance $\Rolr/\Ro$ of the OLR, the 
normalization $\vo$ of the rotation curve, and its shape, parameterized by 
$\beta$. The dependence on $\vo$ is a simple scaling, while that on the other 
three parameters is less trivial. 

We might hope that the dependence on $\Rolr/\Ro$ and $\beta$ is already largely 
described by $\tilde\volr$ defined in \eqn{v-olr}. Indeed, from plotting $\volr$
at fixed bar angle versus $\tilde\volr$, we find that to good accuracy the 
dependence on $\Rolr/\Ro$ and $\beta$ is described by a linear function of the 
form
\beq \label{fit}
	\volr \approx a\,\tilde\volr - (b + c\,\beta)\,\vo.
\eeq
For bar angles $\phi\in[15^\circ,50^\circ]$, \Tab{fit} lists the best-fit values
for $a$, $b$, and $c$ obtained from fitting $\volr$ for $\Rolr/\Ro$ in the range
from 0.8 to 0.95. Note that the values for $b$ and $c$ are always small, i.e.\
$\volr$ is largely given by $a\,\tilde\volr$. The {\sc rms} error made by this
approximation is $0.0035\vo$, while the maximal error is $0.013\vo$
(occuring for $\Rolr/\Ro=0.95$, $\beta=0.2$, $\phi=15^\circ$). 

\placetable{tab:fit}

\ifpreprint
	\begin{table}[t]
	\begin{center} \refstepcounter{table} \label{tab:fit}
			Table~\thetable \\
		Best-fit values for $(a,b,c)$ in \eqn{fit}	
	\begin{tabular}{lccc} 			\\[-2ex] 
				\hline\hline 	\\[-2.5ex]
	$\phi$ 	& $a$ & $b$ & $c$ 		\\
				\hline		\\[-2.3ex]
	15$^\circ$	&1.3549	&0.0761	&0.1362	\\
	20$^\circ$	&1.2686	&0.0642	&0.1120	\\
	25$^\circ$	&1.2003	&0.0526	&0.0892	\\
	30$^\circ$	&1.1424	&0.0406	&0.0711	\\
	35$^\circ$	&1.0895	&0.0298	&0.0538	\\
	40$^\circ$	&1.0420	&0.0200	&0.0423	\\
	45$^\circ$	&1.0012	&0.0103	&0.0316	\\
	50$^\circ$	&0.9653	&0.0012	&0.0238	\\	\hline\hline 
	\end{tabular}
	\end{center}
	\end{table}
\fi	

\ifpreprint	\section{C\fL{omparison with the} O\fL{bserved} \\
			 V\fL{elocity} D\fL{istribution}}
		\label{sec:comp}\noindent
\else		\section{Comparison with the Observed Velocity Distribution}
		\label{sec:comp} \placefigure{fig:fuvobs}
\fi
\Fig{fuvobs} shows the distribution \fuv\ for late-type stars in the solar 
neighborhood (note that the volume sampled by these stars corresponds to the
size of the dot in \fig{olr} referring to a possible solar position). Here, the
velocities $u$ and $v$ are relative to the local standard of rest (LSR) as
measured by Dehnen \& Binney (1998) from a sample of about 14\,000 stars in the
Hipparcos catalogue (ESA 1997). This sample was constructed to be essentially
magnitude limited in order to avoid any kinematic biases. The distribution shown
in \Fig{fuvobs} has been inferred statistically (\cite{deh98}) from the
tangential velocities of about 6000 late-type main-sequence stars ($B-V>0.6\,
$mag) and giants in Dehnen \& Binney's sample.

The distribution \fuv\ has been inferred by maximizing the log-likelihood plus
some penalty function which measures the roughness of $f$ (see Dehnen 1998 for
more details). The latter ensures a smooth distribution and suppresses the
amplification of shot noise. Error estimates using the boot strap method (cf.\
\cite{NR}) indicate that the contours have an uncertainty of about 0.3-3\kms.
The two outermost contours are affected by the way the smoothing is introduced,
and hence are less reliable. All features discussed in this section are of high
significance.
\subsection{The OLR Induced Bi-Modality}
\ifpreprint\noindent\fi
The distribution in \Fig{fuvobs} shows a lot of structures. One of the most
obvious is the bi-modality between a low-velocity component centred on the LSR
motion (solid ellipse) and an intermediate-velocity component at $v\la-30$ and
predominantly negative $u$ (broken ellipse), which contains about a every sixth
late-type star near the Sun. This clear bi-modality was, to the best of my
knowledge, not known from pre-Hipparcos data; the corresponding mean outward
motion of stars with $v\la-30$ was called `$u$-anomaly'. The bi-modality is
hardly present in samples of early-type stars, which almost exclusively populate
the moving groups (sub-peaks) in the low-velocity region (\cite{deh98};
\cite{ch98}; \cite{asi99}; \cite{shc99}).

\ifpreprint
  \begin{figure}[t]
        \centerline{\epsfxsize=85mm\epsfbox[20 17 424 308]{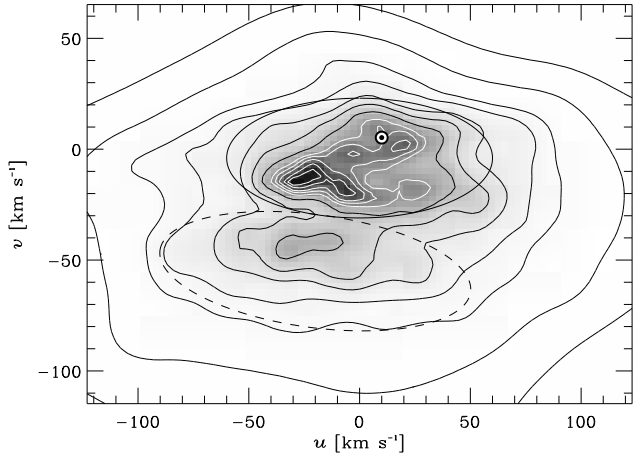}}
        \caption[]{\footnotesize
	The velocity distribution \fuv\ inferred from Hipparcos data for 3\,527
	main-sequence stars with $\bv\ge0.6$ and 2\,491 mainly late-type
	non-main-sequence stars, high-velocity stars excluded, in the solar
	neighborhood (\cite{deh98}). The symbol $\odot$ indicates the solar
	velocity. Samples of early-type stars contribute almost exclusively to
	the low-velocity region ({\em solid ellipse\/}), which contains the most
	prominent moving groups. The region of intermediate velocities ({\em
	broken ellipse\/}) is mainly represented by late-type stars, of which
	$\sim15\%$ fall into this region. Gray scales are linear in $f_0$ and the
	contours contain, from inside out, 2, 6, 12, 21, 33, 50, 68, 80, 90, 95,
	99, and 99.9 percent of all stars. The 1-$\sigma$ uncertainty in the 
	contour lines is between 0.3 and 3\kms.
	\label{fig:fuvobs} }
  \end{figure}
\fi

\subsubsection{Is the Bi-Modality due to the OLR?} \label{sec:alternatives}
\ifpreprint\noindent\fi
The bi-modality present in the locally observed \fuv\ (\fig{fuvobs}) is indeed
very similar to those emerging from the simulations in \Sec{simul}. Moreover,
since we expect, from our previous knowledge of the bar and its estimated
pattern speed, that $\Rolr\sim\Ro$ (cf.\ \Sec{intro}), it is only natural to
identify the observed bi-modality as the OLR feature of the Galactic
bar. However, this may be premature to do and we must first check if there are
viable alternative explanations.

First, can the intermediate-velocity mode at $v\approx-45$ be the relic of a
dispersed stellar (open) cluster or association (the standard explanation for
the formation of moving groups)? The strongest argument against this hypothesis
comes from the age of the participating stars, which from the absence of
early-type stars with intermediate velocities may be inferred to be $\ga8$Gyr
(\cite{deh98}). This high age, corresponding to $\ga40$ orbital times, makes it
very unlikely for any initial moving group to survive Galactic scattering
processes. Moreover, the number of stars in the secondary mode is significantly
higher than that in the low-velocity moving groups, such as the Hyades and
Sirius streams.

Second, could this secondary mode be the result of a merger with a globular
cluster or satellite galaxy? This again can be ruled out, since such a scenario
is highly unlikely to produce a feature with disk-like kinematics (a velocity
deviating from the LSR motion by only $\sim15\%$ instead of an expected 100\%; a
vertical velocity distribution like that of low-velocity stars). Moreover,
Raboud \etal\ (1998) have reported that the $u$-anomaly is caused by
intermediate to high-metallicity stars, while a merger would involve rather
metal poor stars.

Third, could the bi-modality be caused by any other, possibly resonant,
scattering process, e.g.\ due to spiral arm structure? While this possibility is
harder to exclude than the first two, it seems rather unlikely. Spiral structure
should mainly affect stars with epicycles smaller than the inter-arm separation,
while the bi-modality is created by stars with epicycles comparable or larger
than 3\kpc. The possibility of inner Lindblad resonant scattering by some
unknown agent corotating at $\sim25\kpc$, e.g.\ a slowly rotating halo or a
satellite galaxy (note, however, that the magellanic clouds are too far out and
have polar orbits, which disqualifies them as potential agents) was shown by
Weinberg (1994) to be inconsistent with velocity dispersion data.

So, we can conclude that outer Lindblad resonant scattering off the Galactic bar
is presently the only viable explanation that can explain the bi-modality
observed in \fuv\ of late-type disk stars and its absence in early-type stars.

Based on Fux's (1997) $N$-body simulations of the Milky Way, Raboud \etal\
(1998) and Fux (1999b) also attributed the $u$-anomaly to the influence of the
Galactic bar%
\footnote{ 
	These authors, however, did not relate the effect to the OLR, but to the
	fact that the corresponding stars have Jacobi integrals $E_J=E-\Ob L$
	(the Hamiltonian in the corotating frame) that are just high enough to
	enable them to penetrate into the bar region itself. However, there are
	several hints pointing against this interpretation. First, the valley
	between the modes in the simulated $f(\bvel)$ is curved like lines of
	constant energy, while lines of constant $E_J$ are curved the other way
	around. Second, the change in the velocity seperating the modes when
	changing $\Rolr/\Ro$ and $\beta$ cannot be described by a constant value
	of $E_J$ corresponding to the value just allowing penetration into the
	bar. Moreover, the OLR does not occur at constant $E_J$ (cf.\ Figs.\ 2,
	3, and 4 of Fux, 1999b), but roughly at constant $E$.},
but were unable to investigate the parameter space and to achieve a resolution,
both spatially and in velocity, comparable to that in the observed distribution
(a general problem with $N$-body simulations, see my remarks at the beginning of
\Sec{simul:tech}).

\ifpreprint
  \begin{figure}[t]	
	\centerline{\epsfxsize=75mm\epsfbox[22 164 377 688]{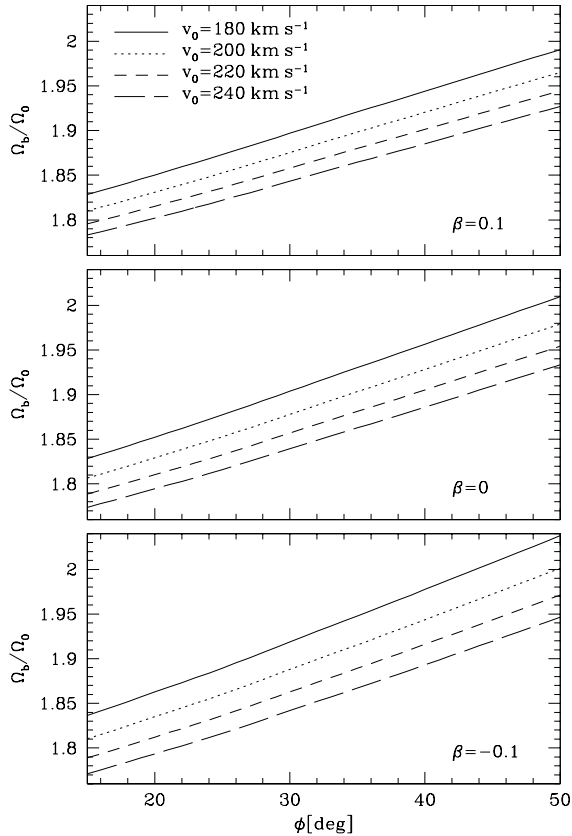}}
	\caption[]{\footnotesize
	The ratio $\Ob/\Oo$ that follows from the new constraint for given
	rotation curve parameters $\vo$, $\beta$, and bar angle $\phi$. The
	typical error in $\Ob/\Oo$ due to the uncertainty in the measured
	$\volr$ is about 1\%, i.e.\ two small ticks on the $y$-axis. A small
	systematic error due to the $v$ motion of the LSR is likely to lead to
	an over-estimation of $\Ob/\Oo$ by up to a few \% for $\phi<45^\circ$
	(see text). \label{fig:ObOz} }
  \end{figure}
\fi
\subsubsection{Implications for the Inner Galaxy} \label{sec:impl}
\ifpreprint \noindent \fi
A quick comparison with the simulations shows that (i) the OLR must be inside
the solar circle, i.e.\ $\Rolr<\Ro$, and (ii) the bar angle $\phi$ in the range
$10^\circ$ to $70^\circ$. We may even go further and make a {\em quantitative}
comparison of the $v$-velocity
\beq \label{new}
	\volr = (-31 \pm 3) \kms,
\eeq
of the valley between LSR and OLR mode with the corresponding velocity measured
in the simulated \fuv. In \Fig{ObOz}, the ratio of the bar's pattern speed over
the local circular frequency that is implied by the new constraint via the
approximation \eqb{fit} is plotted for various possible choices for the local
run of the rotation curve (as given by $\vo$ and $\beta$) and the bar angle
$\phi$. 

\placefigure{fig:ObOz}

The estimate in \fig{ObOz} is subject to some small systematic errors. In
particular, while in the simulations $\volr$ is defined relative to the circular
speed $\vo$ in the absence of any bar, the value \eqb{new} obtained from the
observed \fuv\ has been measured relative to the LSR, which shall deviate from
an exactly circular orbit. From the simulations, it is simple to estimate the
size of this deviation caused by the bar itself (for which it is small,
especially if $\phi\sim45^\circ$), but this is not very meaningful as the LSR,
being defined by low-eccentricity orbits, may well be affected by local spiral
structure, not considered in the simulations. However, even if the azimuthal LSR
motion as large as 10\kms, it would introduce a systematic error of only by 3\%
for $\Ob/\Oo$, i.e.\ an amount that is still smaller than the uncertainty due to
the unknown bar angle $\phi$.

An IAU standard value of $\vo=220\kms$ and a flat rotation curve yield $\Ob$
between 1.8 and 1.9 times $\Oo$. When using the observed proper motion for the
radio source \SgrA, which is thought to be associated with a supermassive black
hole (\cite{eg97}, \cite{ghez98}), to constrain the value of the local circular
frequency, one finds $\Oo\sim28.5\kmskpc$ (\cite{reid99}; \cite{bs99}), and thus
$\Ob\sim51$ to $54\kmskpc$.

See Dehnen (1999c) for a more detailled analysis of the consequences for the
value of $\Ob$, when additionally to the proper motion of \SgrA\ the terminal
gas velocities are used to constrain the run of the rotation curve. The main
result from that study is that the value for $\Ob$ depends only weakly on the
assumed $\Ro$ and $\phi$ (as long as they take reasonable values) and is thus
rather narrowly constraint to be between 50 and 56\kmskpc. Another result from
that study is that corotation of the bar occurs at $\Rcr/\Ro\approx0.5$ to
0.6. This can be compared to the results from stellar- and hydro-dynamical
modeling of the inner Galaxy: Weiner \& Sellwood (1999) and Fux (1999a) report
values in the same range, while Englmaier \& Gerhard (1999) obtain an upper
limit of $0.5$ for $\Rcr/\Ro$, i.e.\ slightly discrepant with the new value.
\subsection{More Similarities with the Simulations}
\ifpreprint\noindent\fi
Apart from the low- vs.\ intermediate-velocity bi-modality, the observed
velocity distribution in \Fig{fuvobs} has also other features in common with the
distributions obtained in the simulations of \Sec{simul}.
\subsubsection{The Structure of the LSR Mode}
\ifpreprint\noindent\fi
The structure of the main mode centred on the LSR, in particular the extension
to $u>0$ at $v\approx-0.1\vo\sim-20\kms$, is at least qualitatively similar,
apart, of course, from the prominent moving groups. In the simulations, this
extension is caused by stars on orbits that are nearly closed and have large
perturbative amplitudes. These orbits originate from near the OLR and therefore
have large deviations from circularity such that they may nonetheless visit
the solar neighborhood. 

However, since these orbits are nearly closed, they are liklely to be affected
by spiral stucture and other local deviations from a smooth force field. As such
effects have not been considered in the simulations of \Sec{simul}, it is not
surprising that simulations and observations do not agree in detail.
\subsubsection{Features due to the outer $1:1$ Resonance}
\ifpreprint\noindent\fi
The ripples at $(u,v)\approx(-80,-5)$ and $(40,5)$ are reminiscent of the
ripples, for instance in \fig{vartime}, caused by the outer 1:1 resonance. This
similarity persists even into details, which is more clearly recognizable when
comparing to simulations with longer integration times, for instance, the middle
panel of \fig{vartime}. Here, the $u<0$ side of the ripple is at smaller $v$
than the $u>0$ side, in agreement with the corresponding features in the
observed \fuv. Moreover, there is a small excess of stars towards larger $v$ at
$|u|<0.4\vo$, which may be compared to the `bump' in the third and fourth
contour (from outside) at $|u|<20\kms$ and $v>20\kms$ in \fig{fuvobs}. This bump
is presumably caused by stars from outside the outer 1:1 resonance.
\ifpreprint	\section{D\fL{iscussion and} C\fL{onclusion}}
		\label{sec:conc} \noindent
\else		\section{Discussion and Conclusion} \label{sec:conc}
\fi
The simulations presented in \Sec{simul} showed how a rotating bar affects the
distribution function of a surrounding stellar disk. Since the non-axisymmetric
component of the bar-induced forces falls off steeply with radius ($\propto
r^{-4}$ for the quadrupole) reaching about 1\% of the total force at $\Ro$,
the influence of the bar is restricted to orbits which are nearly in resonance
with it. This influence is strongest for the outer Lindblad resonance (OLR),
which occurs for orbits whose radial and azimuthal frequencies obey the relation
$\Ob=\op+\half\oR$, where $\Ob$ is the bar's rotation rate (pattern speed). At
orbits that are nearly in OLR, an otherwise smooth stellar distribution function
becomes distorted. At any given position in the disk, this distortion is visible
in the observable stellar velocity distribution $f(u,v)$ if near-resonant orbits
(i) pass through this position and (ii) are sufficiently populated. For a
distribution function designed to resemble the old stellar disk of the Milky
Way, the OLR leaves its imprint in the velocity distribution over a wide range
of possible position in the disk.

For positions $(\Ro,\phi$) orientated relative to the bar at angles
$\phi\in[10^\circ,70^\circ]$ and situated outside the radius $\Rolr$ where
circular orbits are in OLR, the resulting feature in $f(u,v)$ is a clear
bi-modality: apart from the dominant stellar component with velocities similar
to the local standard of rest (LSR), there is a secondary mode at $u,v<0$. This
OLR mode consists of stars with mean outward motion and slower rotation
velocities than the LSR. For $\Ro<\Rolr$, the OLR induced feature occurs at
higher rotation velocities than the LSR and no longer takes the form of a clear
bi-modality. As has been demonstrated in \Sec{orbs}, all these changes in the
behaviour of \fuv\ can be well understood from the properties of the closed
orbits in the outer parts of a barred galaxy.

Extensive simulations showed that the precise velocity of the OLR-induced
feature depends mainly on four parameters: the normalization and shape of the
underlying circular speed curve, $\vc(R)$, the observer's relative position with
respect to the OLR, $\Rolr/\Ro$, and their orientation relative to the bar,
measured by the bar angle $\phi$. Note that the dependence on $\vc(R)$ is
restricted to its run in the region visited by near-resonant slightly eccentric
orbits, i.e.\ between about $0.7\Ro$ and $\Ro$ for $\Rolr=0.9\Ro$. These
dependences are such that the (negative) $v$-velocity, $\volr$, separating the
OLR mode from the LSR mode of $f_0$, increases (becomes smaller in modulus) with
increasing bar angle $\phi$, decreasing ratio $\Rolr/\Ro$, $\vo\equiv\vc(\Ro)$,
and $\beta\equiv\D\ln\vc/\D\ln R$.

The velocity distribution $f(u,v)$ actually observed for late-type stars in the
solar neighborhood (\fig{fuvobs}) shows indeed a secondary mode at $u,v<0$,
which is very similar to the OLR modes emerging in the simulations. As I have
argued in \Sec{alternatives}, there exists no satisfying explanation of this
seeming anomaly other than being the feature induced by the OLR of the Galactic
bar. Interestingly, the observed \fuv\ has also other features in common with
the simulated distributions. Firstly, extensions towards high $u$ (inward
moving) at slightly negative $v$ may be related to a ridge in \fig{fuvobs} at
$v\approx{-}20\kms$ connecting the Hyades and Pleiades stream and reaching at
least to $u\approx60\kms$. Secondly, ripples due to the outer 1:1 resonance seen
in the simulated $f(u,v)$ at $v\ge0$ and large $|u|$ may be related to the
features in \fig{fuvobs} at $(u,v)=(-80,-5)$ and $(40,5)\kms$.

First of all, this means that the structure of the local velocity distrubution
for late-type stars provides a clear evidence for the existence of the Galactic
bar, completely independent of any observation of the inner Galaxy itself.
Moreover, this structure implies that the Sun is situated slightly outside of
the radius $\Rolr$ and at bar angles $\phi$ between about $10^\circ$ and
$70^\circ$. One may even use the observed value, $\volr$, for the $v$-velocity
seperating the OLR and LSR mode in order to deduce that the ratio between the
bar's pattern speed $\Ob$ and the local circular speed $\Oo$ is between 1.75 and
2. The uncertainty here is dominated by the uncertainty in the bar angle $\phi$,
which according to other evidence must lie somewhere in the range 20 to 45
degrees. Combining this result with the value $\Oo\approx28.5\kmskpc$ derived from
the proper motion of the radio source \SgrA\ at the Galactic centre
(\cite{bs99}; \cite{reid99}) yields $\Ob=(53\pm3)\kmskpc$. 

Clearly, the conclusion on the precise value of $\Ob$ drawn on the basis of the
simulations of \Sec{simul} is subject to (presumably small) systematic errors,
originating from the simplifications made in the simulations and from the
unknown $v$ motion of the LSR. To improve on this, one needs to (i) use a more
realistic model for the Galactic potential, (ii) compare not only $\volr$ of the
simulated \fuv\ but also the feature of the 1:1 resonance (to reduce systematic
errors due to the unknown LSR azimuthal motion), and (iii) combine this with
hydro-dynamical simulations, which can be compared directly to the gas
velocities observed in the entire inner Galaxy, constraining both the bar
parameters and the rotation curve. Such simulations, in conjunction with the
constraint set by the proper motion of \SgrA, are likely to determine the
pattern speed, orientation, and structure of the bar as well as the rotation
curve and hence mass distribution of the inner Galaxy. In such an analysis, the
constraints set by the local \fuv\ constitute important new ingredients as
they largely reduce the freedom for the pattern speed $\Ob$, which is not very
well constraint by modeling of the inner Galaxy as is apparent from the 
diverging results obtained in the past by different modellers.

Thus, the velocity distribution observed locally imposes important new
constraints for the structure of the inner Galaxy independent of and
complementary to the photometry and kinematics, both stellar and gaseous, of
that region itself.
%
%
\acknowledgements
I am grateful to James Binney and to the anonymous referee for useful comments.
This work was financially supported by grants from PPARC and the Max-Planck
Gesellschaft.
%
%
\ifpreprint \relax \else
  \clearpage
  \begin{deluxetable}{llr@{}c@{}l}
	\tablewidth{0pt}
	\tablecaption{Default values for the simulation parameters
			\label{tab:default}}
	\tablehead{\multicolumn{5}{c}{parameters to be varied 
			\hspace{1cm}default value} }
	\tablecolumns{5}
	\startdata
	shape of rotation curve & $\beta$       & 0  &           &      \\
	bar angle               & $\phi$        & 25 &&$\!\!{}^\circ$   \\
	position w.r.t.\ OLR    & $\Rolr/\Ro$   & 0  &. &9              \\
	bar strength            & $\alpha$      & 0  &. &01             \\
	integration time        & $t_2$         & 4  &  &$T_{\rm b}$    \\
	\cutinhead{parameters kept fixed \hspace{1.2cm}default value }
	bar size		& $\Rb/\Rcr$	& 0  &. &8		\\
	disk scale length	& $R_s/\Ro$	& 0  &. &33		\\
	local velocity dispersion & $\sR(\Ro)/\vo$ & 0 &. &2	\\
	$\sR$ scale length	& $R_\sigma/\Ro$& 1  &	&		\\
	bar growth time		& $t_1$		& 0  &. &5 $t_2$	\\
	\enddata
  \end{deluxetable}
  \begin{deluxetable}{l@{$\qquad$}r@{.}l@{$\qquad$}r@{.}l@{$\qquad$}r@{.}l}
	\tablewidth{0pt}
	\tablecaption{Best-fit values for $(a,b,c)$ in \eqn{fit}\label{tab:fit}}
	\tablehead{$\phi$ & \multicolumn{2}{c}{$a$} &
		\multicolumn{2}{c}{$b$} & \multicolumn{2}{c}{$c$}}
	\startdata
	15$^\circ$	&1&3549	&0&0761	&0&1362	\\
	20$^\circ$	&1&2686	&0&0642	&0&1120	\\
	25$^\circ$	&1&2003	&0&0526	&0&0892	\\
	30$^\circ$	&1&1424	&0&0406	&0&0711	\\
	35$^\circ$	&1&0895	&0&0298	&0&0538	\\
	40$^\circ$	&1&0420	&0&0200	&0&0423	\\
	45$^\circ$	&1&0012	&0&0103	&0&0316	\\
	50$^\circ$	&0&9653	&0&0012	&0&0238	\\
	\enddata
  \end{deluxetable}
\fi	

%
%

\ifpreprint
  \def\thebibliography#1{\subsection*{R\fL{eferences}}
    \list{\null}{\leftmargin 1.2em\labelwidth0pt\labelsep0pt\itemindent -1.2em
    \itemsep0pt plus 0.1pt
    \parsep0pt plus 0.1pt
    \parskip0pt plus 0.1pt
    \usecounter{enumi}}
    \def\refpar{\relax}
    \def\newblock{\hskip .11em plus .33em minus .07em}
    \sloppy\clubpenalty4000\widowpenalty4000
    \sfcode`\.=1000\relax
    \footnotesize}
  \def\endthebibliography{\endlist}
\fi

%
%

\ifpreprint \relax \else
\clearpage \onecolumn

\begin{figure}\caption[]{
	Closed orbits ({\it solid\/}) just inside and outside the OLR of a
	rotating central bar ({\it shaded ellipse\/}). The circles ({\it
	dotted\/}) depict the positions of the ILR, CR, and OLR (from inside out)
	for circular orbits. Note the change of the orbits' orientation at the
	OLR, resulting in the crossing of closed orbits at four azimuths.  A
	possible position of the Sun is shown as filled circle. The bar angle
	$\phi$ is indicated for the case of a clockwise rotating bar.
	\label{fig:olr} }
\end{figure}

\begin{figure}\caption[]{
	Simulated velocity distributions \fuv: variation with bar angle.
	The remaining parameters are fixed at their default values in
	\Tab{default}. \label{fig:varphi} }
\end{figure}

\begin{figure}\caption[]{
	Simulated velocity distributions \fuv: variation with 
	integration time $t_2$ ($T_b$ denotes the bar rotation period).
	The bar-growth time is set to $t_1=0.5t_2$.
	\label{fig:vartime} }
\end{figure}

\begin{figure}\caption[]{
	Simulated velocity distributions \fuv: variation with $\Rolr/\Ro$. 
	\label{fig:varpos} }
\end{figure}

\begin{figure}\caption[]{
	Simulated velocity distributions \fuv: variation with $\beta$,
	parameterizing the shape of the rotation curve. \label{fig:varbeta} }
\end{figure}

\begin{figure}\caption[]{
	Simulated velocity distributions \fuv: variation with bar 
	strength, parameterized by $\alpha$. \label{fig:varalpha} }
\end{figure}

\begin{figure}\caption[]{
		Characteristic diagram (plot of $R$ vs.\ $L$ on the bar's minor
		axis) of the stable ({\em solid}) and unstable ({\em dotted})
		closed orbits in the outer parts of the model with default
		settings. Orbits corresponding to high-order resonances ($n<-2$)
		are not displayed. The dashed line corresponds to the circular
		orbits in the unbarred model with the square indicating the
		OLR. \label{fig:char}}
\end{figure}

\begin{figure}\caption[]{
		Closed orbits of the families also displayed in \fig{char}.  The
		OLR is indicated by the bold circle. Solid and dotted curves
		refer to stable and unstable orbits, respectively.  The x$_1$(1)
		orbits extend to larger radii, while for the other four orbital
		families the innermost and outermost orbit is shown (note the
		difference in scale). The 1:1 resonant orbits (d \& e) also
		exist as reflected versions, such that each family as a whole is
		bi-symmetric.  \label{fig:orbits}}
\end{figure}

\begin{figure}\caption[]{
		The velocity distribution \fuv\ inferred from Hipparcos data for
		3\,527 main-sequence stars with $\bv\ge0.6$ and 2\,491 mainly
		late-type non-main-sequence stars, high-velocity stars excluded,
		in the solar neighborhood (\cite{deh98}). The symbol $\odot$
		indicates the solar velocity. Samples of early-type stars
		contribute almost exclusively to the low-velocity region ({\em
		solid ellipse\/}), which contains the most prominent moving
		groups. The region of intermediate velocities ({\em broken
		ellipse\/}) is mainly represented by late-type stars, of which
		$\sim15\%$ fall into this region. Gray scales are linear in
		$f_0$ and the contours contain, from inside out, 2, 6, 12, 21,
		33, 50, 68, 80, 90, 95, 99, and 99.9 percent of all stars. The
		1-$\sigma$ uncertainty in the contour lines is between 0.3 and
		3\kms.  \label{fig:fuvobs} }
\end{figure}

\begin{figure}\caption[]{
		The ratio $\Ob/\Oo$ that follows from the new constraint for
		given rotation curve parameters $\vo$, $\beta$, and bar angle
		$\phi$. The typical error in $\Ob/\Oo$ due to the uncertainty in
		the measured $\volr$ is about 1\%, i.e.\ two small ticks on the
		$y$-axis. A small systematic error due to the $v$ motion of the
		LSR is likely to lead to an over-estimation of $\Ob/\Oo$ by up
		to a few \% for $\phi<45^\circ$ (see text). \label{fig:ObOz} }
\end{figure}
\vfill

%
%

	\clearpage
	\centerline{\epsfxsize=100mm\epsfbox[90 220 550 676]{Dehnen.FIG1.ps}}
        \centerline{\fig{olr}}
	\centerline{\epsfxsize=80mm\epsfbox[6 4 252 634]{Dehnen.FIG2.ps}}
        \centerline{\fig{varphi}}
	\clearpage
	\centerline{\epsfxsize=60mm\epsfbox[7 6 169 274]{Dehnen.FIG3.ps}}
        \centerline{\fig{vartime}}
	\clearpage
	\centerline{\epsfxsize=60mm\epsfbox[6 19 159 411]{Dehnen.FIG4.ps}}
        \centerline{\fig{varpos}}
	\clearpage
	\centerline{\epsfxsize=60mm\epsfbox[7 0 166 425]{Dehnen.FIG5.ps}}
        \centerline{\fig{varbeta}}
	\clearpage
	\centerline{\epsfxsize=60mm\epsfbox[7 6 169 274]{Dehnen.FIG6.ps}}
        \centerline{\fig{varalpha}}
	\clearpage
	\centerline{\epsfxsize=90mm\epsfbox[28 166 410 706]{Dehnen.FIG7.ps}}
        \centerline{\fig{char}}
	\clearpage
	\centerline{\epsfxsize=120mm\epsfbox[238 204 560 688]{Dehnen.FIG8.ps}}
        \centerline{\fig{orbits}}
	\clearpage
        \centerline{\epsfxsize=120mm\epsfbox[20 17 424 308]{Dehnen.FIG9.ps}}
        \centerline{\fig{fuvobs}}
	\clearpage
	\centerline{\epsfxsize=110mm\epsfbox[22 164 377 688]{Dehnen.FIG10.ps}}
        \centerline{\fig{ObOz}}

\fi 

\end{document}